\documentclass{aa}
\usepackage{graphicx,natbib}
\bibpunct{(}{)}{;}{a}{}{,} 
\def\deg{$^\circ$}

\begin{document}
\sloppypar

\title{INTEGRAL observations of the cosmic X-ray background in
   the 5-100 keV range via occultation by the Earth}

\author{E.Churazov$^{1,2}$,  R.Sunyaev$^{1,2}$, M.Revnivtsev$^{1,2}$,
S.Sazonov$^{1,2}$, S.Molkov$^{1,2}$, S. Grebenev$^{1}$,
C.Winkler$^{3}$, A.Parmar$^{3}$, A. Bazzano$^{4}$, M. Falanga$^{5}$,
A. Gros$^{5}$, F.Lebrun$^{5,6}$, L. Natalucci$^{4}$,
P. Ubertini$^{4}$, J.-P.Roques$^{7}$, L.Bouchet$^{7}$,
E.Jourdain$^{7}$, J.Knoedlseder$^{7}$, R.Diehl$^{8}$,
C.Budtz-Jorgensen$^{9}$, S.Brandt$^{9}$, N.Lund$^{9}$,
N.J.Westergaard$^{9}$, A.Neronov$^{10}$, M.T\"urler$^{10}$,
M.Chernyakova$^{10}$, R.Walter$^{10}$, N.Produit$^{10}$,
N.Mowlavi$^{10}$, J.M.Mas-Hesse$^{11}$, A. Domingo$^{12}$,
N.Gehrels$^{13}$, E.Kuulkers$^{14}$, P.Kretschmar$^{14}$,
M.Schmidt$^{15}$}


\institute{
Space Research Institute, Russian Academy of
 Sciences, Profsoyuznaya 84/32, 117997 Moscow, Russia
\and 
Max-Planck-Institut f\"{u}r Astrophysik,
         Karl-Schwarzschild-Str. 1, 85740 Garching bei M\"{u}nchen,
           Germany 
\and
        ESA, ESTEC (SCI-SA), Keplerlaan 1, NL-2201, AZ Noordwijk, 
The Netherlands 
\and
IASF/INAF, Via Fosso Del Cavaliere 100, I-00133 Rome, Italy
\and
Service d'Astrophysique, DAPNIA/DSM/CEA, 91191 Gif-sur-Yvette, France
\and
APC-UMR 7164, 11 place M. Berthelot, 75231 Paris, France
\and
 Centre d'Etude Spatiale des Rayonnements
9, avenue du Colonel Roche, 31028 Toulouse Cedex 4, France
\and
Max-Planck-Institut fuer extraterrestrische Physik,
Giessenbachstr., 
85748 Garching,
Germany
\and
Danish National Space Center, 
Juliane Maries Vej 30, 
2100 Copenhagen, 
Denmark 
\and
INTEGRAL Science Data Centre,
Chemin d'Ecogia 16,
CH-1290 Versoix,
Switzerland
 \and
Centro de Astrobiolog\'{\i}a (CSIC-INTA), POB 50727, 28080 Madrid, Spain
\and
Laboratorio de Astrof\'{\i}sica Espacial y F\'{\i}sica Fundamental (LAEFF-INTA),
POB 50727, 28080 Madrid
\and
Goddard Space Flight Center, Bldg. 2, Room 245, Code 661.0, Greenbelt, MD 20771, USA
\and
ISOC, ESAC/ESA, Apartado 50727, 28080 Madrid, Spain 
\and
ESA-ESOC, Mission Operations Department, Robert-Bosch Str. 5, 64293 Darmstadt, Germany
}

\date{}

\authorrunning{Churazov et al.}
\titlerunning{INTEGRAL observations of the Earth/CXB}


\abstract{}{
We study the spectrum of the cosmic X-ray background (CXB) in energy
range $\sim$5-100 keV.}{
Early in 2006 the INTEGRAL observatory performed a series of four
30ksec observations with the Earth disk crossing the field of view of
the instruments.  The modulation of the aperture flux due to
occultation of extragalactic objects by the Earth disk was used to
obtain the spectrum of the Cosmic X-ray Background(CXB).  Various
sources of contamination were evaluated, including compact sources,
Galactic Ridge emission, CXB reflection by the Earth atmosphere,
cosmic ray induced emission by the Earth atmosphere and the Earth
auroral emission.}  {The spectrum of the cosmic X-ray background in
the energy band 5-100 keV is obtained.  The shape of the spectrum is
consistent with that obtained previously by the HEAO-1 observatory, while
the normalization is $\sim$10\% higher.  This difference in
normalization can  (at least partly) be traced to the different
assumptions on the absolute flux from the Crab Nebulae.  
  The increase relative to the earlier adopted value of
the absolute flux of the CXB near the energy of maximum luminosity
(20-50 keV) has direct implications for the energy release of
supermassive black holes in the Universe and their growth at the epoch
of the CXB origin.}{}{}
\maketitle

\section{Introduction}
It is well established \cite[e.g.][]{giacconi01} that the bulk of the CXB
emission below $\sim$5 keV is provided by the numerous active galactic
nuclei (AGN) - accreting supermassive black holes, which span large
range of redshifts from 0 up to 6. At these low energies X-ray
mirror telescopes can directly resolve and count discrete sources. The
resolved fraction drops from $\approx$80\% at 2-6 keV to 50-70\% at 6-10 keV
(see Brandt \& Hasinger, 2005 for review).

At energies above 10 keV the efficiency of X-ray
mirrors declines and at present it is impossible to resolve more than
few percent of the CXB emission in this regime. It is believed that
AGN still dominate CXB at higher energies (at least up to hundreds of
keV), although the extrapolation of the low energy data is complicated
by the presence of several distinct populations of AGN with different
spectra and intrinsic absorption column densities (Sy I, Sy II and
quasars, blazars, etc., see e.g. \citealt{setti89,
comastri95,zdziarski96}). At the same time the peak of
the CXB luminosity is around 30 keV and accurate measurements of the
CXB flux at high energies (even if we can not resolve individual
objects) are important to understand the energy release in the Universe
and the contribution of various types of objects to it. 

Such measurements are complicated because instruments working in the
energy range from tens to hundreds keV 
are often dominated by the internal detector background, caused by
the interactions of charged particles with the detector material. To
decompose the total background into particle-induced background and the CXB
contribution one needs either a very good model of the internal detector
background or two observations having different relative contributions
of these two components. The latter approach was behind the INTEGRAL
observations of the Earth which uses the Earth disk as a natural
screen to modulate the CXB flux coming on to the detectors.

A similar approach has already been used for the same purpose in the
analysis of early space X-ray experiments. In particular the HEAO-1
observatory used a movable 5 cm thick CsI crystal to partly block the
instrument field of view and to modulate the CXB signal (Kinzer et
al. 1997, Gruber et al. 1999). Here we report the results of the
first INTEGRAL observations of the Earth performed in 2006. During
these observations the Earth was drifting through the field of view of
the INTEGRAL instruments producing a modulation of the flux with an
amplitude of the order of 200 mCrab at 30 keV.

The structure of the paper is as follows. In Section 2 we present the
details of the Earth observations with INTEGRAL in 2006. In Section 3
we introduce various components contributing to the light curves
recorded by the different instruments. In Section 4 we describe the CXB
spectrum derived from the INTEGRAL data. The last section summarizes
our findings.

\section{Observations}
{\em INTEGRAL} (The {\bf Inte}rnational {\bf G}amma-{\bf R}ay {\bf
A}strophysics {\bf L}aboratory; Winkler et al.\ 2003) is an ESA
scientific mission dedicated to fine spectroscopy and imaging of
celestial $\gamma$-ray sources in the energy range 15\,keV to
10\,MeV. 

The primary imaging instrument onboard {\em INTEGRAL} is IBIS (Ubertini
et al. 2003) -- a coded-mask aperture telescope with the CdTe-based
detector ISGRI (Lebrun et al. 2003). It has a high sensitivity in the
$20-200$~keV range and has a spatial resolution of better than 10$'$.

 The best energy resolution (from $\sim$1.5 to 2.2 keV for
energies in the 50-1000 keV range) is provided by the SPI telescope -
a coded mask germanium spectrometer consisting of 19 individual Ge
detectors(Vedrenne et al., 2003, Attie et al., 2003).

 In addition {\em INTEGRAL} provides simultaneous monitoring of sources
in the X-ray (3--35\,keV; JEM-X, see Lund et al. 2003) and optical
(V-band, 550\,nm; OMC, see Mas-Hesse et al., 2003) energy ranges.

The observations used in the analysis were made in January-February
2006 in four separate periods (Table \ref{tab:obslog}). Each observation
lasted about 30 ksec.

\begin{table} 
\caption{The list of Earth observations by INTEGRAL. The pointing
direction corresponds to the beginning of the observation and is
drifting slowly (by $\sim$10' during each observation). Each observation
lasted about 30 ksec.}
\begin{tabular}{l l l l}
\hline
Revolution & Start Date (UT) & Pointing \\
 &   & $\alpha,\delta$, deg,  J2000  \\
\hline
\hline
401 & 2006-01-24  & 252.0~~-60.8\\
404 & 2006-02-02  & 251.4~~-61.3\\
405 & 2006-02-05  & 251.1~~-61.1\\
406 & 2006-02-08  & 250.9~~-60.9\\
\hline
\hline
\end{tabular}
\label{tab:obslog}
\end{table}

\subsection{Observing strategy}
\begin{figure}
\includegraphics[width=\columnwidth]{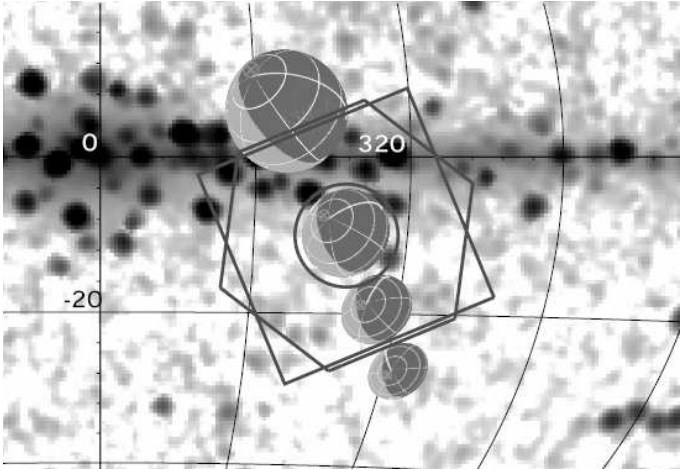}
\caption{Illustration of the INTEGRAL Earth observing mode. Zero
sensitivity FoVs of JEM-X, IBIS and SPI are shown with a circle, box
and hexagon respectively superposed on to the RXTE 3-20 keV slew
map. In this map many compact sources and extended X-ray emission
associated with the Galactic Ridge are visible. In the course of the
observation the pointing direction of the telescopes remains the same,
while the Earth crosses the instruments FoVs.  Day side of the Earth
is shown by lighter shade of gray. The linear sizes of the fully coded
FoVs are roughly twice smaller. In the course of the observation the
Earth moves from positive to negative latitudes. The distance from the
Earth increases rapidly during this part of the 3-day INTEGRAL orbit
and the angular size of the Earth disk decreases.
\label{fig:obsmode}
}
\end{figure}

The standard observing mode of INTEGRAL requires the Earth limb to be
at least 15\deg away from the main axis. This separation is required
for the observatory star tracker to operate in a normal way and
provide information needed for 3-axis stabilization of the
spacecraft. These 15 degrees approximately correspond to the radius of
the SPI and IBIS field of views (at zero sensitivity). In particular,
SPI has a hexagonal field of view (see Fig.\ref{fig:obsmode}) with a
side to side angular size at zero sensitivity of 30.5\deg. The IBIS
field of view is a rectangle having 29\deg on side at zero
sensitivity. JEM-X has the smallest field of view with a 13.2\deg
diameter at zero response. A special operational procedure was
developed by  the INTEGRAL Science Operations Centre (ISOC) and the
INTEGRAL Mission Operations Centre (MOC), in consultation with the
instruments teams, in order to on one hand allow the Earth to be
within the FoVs of the instruments and on the other hand to ensure a
safe mode of operations. The aim was to maximize the solid angle
within the FoVs subtended by the Earth and to have sufficiently long
observations.

All observations were performed during the rising part of the 3 days
satellite orbit, a few hours after the perigee passage. The Earth
center (as seen from the satellite) is making a certain track on the
celestial sphere. As a first step (with the star tracker on) the main
axis of the satellite was pointed towards the position where the Earth
would be $\sim$6 hours after perigee exit. The satellite was then in a
controlled 3-axis stabilization while the Earth was drifting towards
this point. When the Earth limb came within 15\deg from the main axis
the star tracker was switched off. Once the Earth crossed the INTEGRAL
FoV and the distance between the satellite axis and the Earth limb
became larger than 15\deg$~$ the star tracker was switched back-on
restoring the controlled 3-axis stabilization. During the period when
the star tracker was off the satellite was passively drifting. The
total amplitude of the drift was about 10' and interpolation of the
satellite attitude information before and after the drift allows
reconstruction of the satellite orientation at any moment with an
accuracy better than 10''.  The distance from the Earth during this
period varied from $\sim$40,000 to $\sim$100,000 km. When the Earth
center was close to the main axis of the satellite the angular size of
the Earth (radius) was $\sim$5.4\deg, corresponding to a subtended
solid angle of $\sim$90 sq.deg. As is further discussed in section
\ref{sec:method} (see also Fig.\ref{fig:components} and
\ref{fig:lcurves}) the modulation of the CXB flux by the Earth disk is
the main source of the flux variations observed by JEM-X, IBIS and
SPI.

Schematically this mode of observations is illustrated in
Fig.\ref{fig:obsmode}. The FoVs of JEM-X, IBIS and SPI are shown with
a circle, box and hexagon respectively superposed on to the RXTE 3-20
keV sky map (Revnivtsev et al., 2004). Compact  sources in this
map show up as  dark patches, while the Galactic Ridge emission is
visible as a grey  strip along the Galactic plane. The Earth position
is schematically shown for 4 successive instants, separated by $\sim$
8.1, 9.3 and 10.8 ksec respectively. The day and night
sides of the Earth are indicated by the lighter and darker shades of
grey respectively.

\subsection{``Background'' field}
Ideally one would like to observe the Earth modulated CXB signal
having an ``empty'' (extragalactic) field as a background. However due
to the requirement of observing the Earth at the beginning of a
revolution and the properties of the INTEGRAL orbit the pointing
direction of the satellite was set to $l\sim 327$, $b\sim-10$,
i.e. rather close to the Galactic plane (Fig.\ref{fig:obsmode}). As a
result the recorded variations of the count rates were not only due to
the CXB modulation but also due to occultation of compact sources and
the Galactic Ridge emission by the Earth disk. This is further
discussed in section \ref{sec:compact}).

\begin{figure*}
\includegraphics[width=\textwidth]{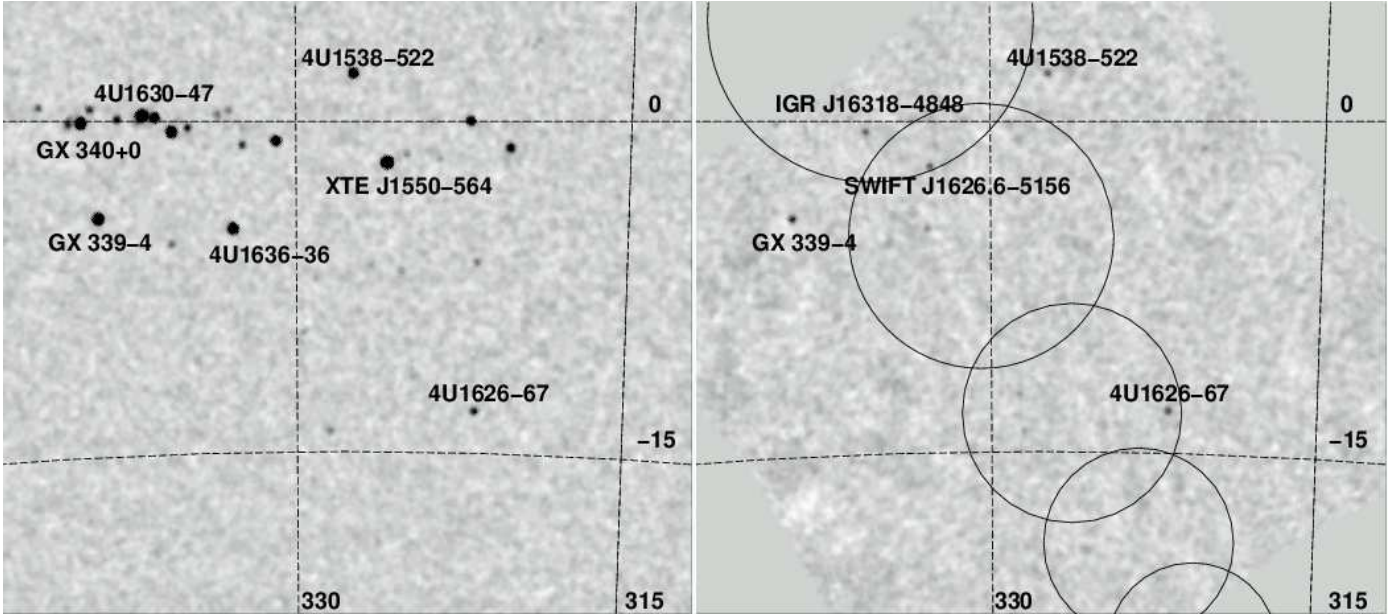}
\caption{``Background'' field: map built using the data averaged over
several years (left) and actual image during Earth observations
(right). The circles shows the Earth disk during several consecutive
moments of time. Darker shades of grey correspond to higher S/N ratio.  
\label{fig:image_isgri}
}
\end{figure*}

The same field has been observed by INTEGRAL multiple times during the
regular observational program. Fig.\ref{fig:image_isgri} shows the
17-60 keV image averaged over multiple observations during several
years of INTEGRAL operations (left) and the much less deep image obtained
by averaging 4 observations during the Earth observations (right).
The list of sources detected with S/N $>4.5\sigma$ during the Earth
observation is given in Table \ref{tab:src}. One of the sources - IGR
J17062-6143 - was found during the Earth observations. The source is
apparently a transient, since it is not present in the images
averaged over all previous observations.  The last column in the table
indicates whether the source was obscured by the Earth disk in the course of
observations. Further discussion on the contamination of the CXB
signal by compact sources is given in section \ref{sec:compact}.

\begin{table}
\caption{The list of compact sources detected during the Earth
observations with INTEGRAL. The flux in the 17-60 keV band is
quoted. In this band 1 mCrab corresponds to $\sim 1.4~10^{-11}~{\rm ergs~cm^{-2}~s^{-1}}.$ The new source IGR J17062-6143 was discovered during the Earth
observations with INTEGRAL.}
\begin{tabular}{l|c|c|c}
Source&Flux,mCrab&S/N&Occult.\\
\hline
GX339-4           & $30.9 \pm  1.4$ &  21.6 & N\\
4U1626-67         & $12.0 \pm   0.7$ &  16.3 & Y\\
SWIFT J1626.6-5156& $13.9 \pm   1.0$ &  14.3 & Y\\
4U 1538-522       & $17.8 \pm   1.3$ &  13.5 & Y\\
IGR J16318-4848   & $20.9 \pm   1.6$ &  13.0 & Y\\
4U 1636-536       & $ 6.7 \pm   0.8$ &   8.5 & Y\\
PSR 1509-58       & $ 7.2 \pm   1.2$ &   6.1 & N\\
IGR J17062-6143$^*$&$  3.4\pm   0.6$ &   5.8 & Y\\
GX 340+0          & $60.2 \pm  10.4$ &   5.8 & Y\\
NGC 6300          & $ 3.3 \pm   0.6$ &   5.4 & Y\\
XTE J1701-462     & $15.5 \pm   3.3$ &   4.7 & N\\
\hline
\end{tabular}
\begin{list}{}
\item $^*$ - newly detected source
\end{list}
\label{tab:src}
\end{table}

\subsection{JEM-X}
\label{sec:jem}
Initial reduction of JEM-X data was done using the standard INTEGRAL
Off-line Science Analysis software version 5.1 (OSA-5.1) distributed
by the INTEGRAL Science Data Centre.  We use the event lists to
which, an arrival time, energy gain and position gain corrections have
been applied (the so called COR level data in the OSA-5.1 notations).
Using these lists the light curves and spectra for the whole detector
(i.e. ignoring position information) have been generated. To convert
the detector count rates into photon rates we used the Crab nebula
observations in revolution number 300, when the Crab was in the center
of JEM-X FoV (see more discussion on the cross calibration in
Sect. \ref{sect:crab}).  For the Earth observations the effective area
was calculated over the part of the disk within JEM-X FoV, taking into
account position dependent vignetting. In the subsequent analysis we
used only the data from JEM-X unit 1 which has a more accurate
calibration than unit 2.

Each of the Earth observations started immediately after switching on
the spacecraft instruments at the Earth radiation belts exit. The
first 4 ksec of each observation were discarded because of the strong
variations of the JEM-X gain  usually accompanying the instrument
turn-on.

\begin{figure}
\includegraphics[width=\columnwidth]{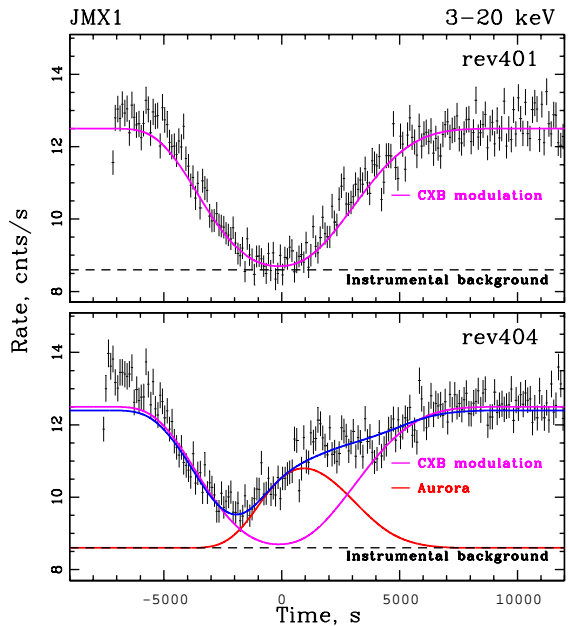}
\caption{Count rates collected by JEM-X1 detector in the 3-20 energy
band during two sets of Earth observations carried out on Jan. 24-25
and Feb. 2-3, 2006 (revolutions 401 and 404 respectively).  Solid thick lines
show the contribution of CXB flux modulation and the Earth Auroral
emission. The auroral model assumes that the emission comes from the
circular region with a radius of order of $0.1\times R_{E}$ around the
Earth North magnetic pole. Such model is probably too simple to rely
upon and the revolutions 404 and 405 (showing clear signs of auroral
emission) have been removed from the analysis.
Short dashed line shows the instrumental background
level. Zero time in these plots corresponds to maximum coverage of the
JEM-X FoV by the Earth disk.
\label{fig:aurora}
}
\end{figure}

 Light curves of the JEM-X detector during two observations
(revolutions 401 and 406) can be well described in terms of our
CXB-modulation model. In other two observations (revolutions 404 and
405) an additional component is clearly present in the detector light
curve, which can be interpreted as due to auroral emission from the
Earth (see section \ref{sec:aurora}). The examples of JEM-X light
curves (with and without evidence for auroral emission) are shown in
Fig.\ref{fig:aurora}. For the CXB analysis we used only the JEM-X data
obtained during revolutions 401 and 406 where the contribution from
the auroral emission is small (see section \ref{sec:aurora}).

\subsection{IBIS}
For the IBIS/ISGRI data the energies of the events were calculated
using the standard conversion tables available in the OSA 5.1
distribution. The secular evolution of the detector gain was corrected
using the position of fluorescent line of tungsten at $\sim$60 keV in
the detector background spectrum. The count rate for the whole
detector in 1 keV wide channels was used for subsequent analysis.

\subsection{SPI}
SPI  spectra in  0.5 keV wide energy bins were accumulated
for 200 second intervals from standard energy-calibrated events. Over
this time interval, the sky as occulted by the Earth can be assumed to
be constant. The light curves measured by each of the 17 detectors of
SPI were used {\bf (independently)} in the subsequent analysis.

\subsection{OMC}
As explained in Sect. 2.1, the star trackers of the spacecraft were
switched off when the Earth limb was at less than 15\deg  from the
main axis, so that the satellite was passively drifting during the
occultation. The Optical Monitoring Camera (OMC) was configured to
obtain several reference images of the stellar background before and
after the Earth occultation, which could be used as a backup to
recover the attitude of the spacecraft if there were any technical
problem. OMC obtained a sequence of 13 images of $512\times 512$
pixels (corresponding to 2.5\deg $\times$ 2.5\deg), with integration
time of 10 s. The drift measured on the OMC images during the first
observation (revolution 401) was $\Delta Y_{sc} = -5.5'$ and
$\Delta Z_{sc} = +10.9'$, totalling 12.2' in a period
of time $\Delta T = 06h~48m~33s$. $Y_{sc}$ and $Z_{sc}$ are the
spacecraft reference axes. This drift is fully consistent with
the interpolation of the spacecraft attitude control system data,
used to determine the pointing direction. The diffuse light originated by the
illuminated fraction of the Earth was detectable by OMC already 1 hour
before entering the Earth limb (when the OMC axis was at around
18\deg from the Earth centre). Although the integration time was
reduced to 1 s when the Earth was within the OMC field of view, the
diffuse light was bright enough to completely saturate the OMC CCD.

\section{Methods and analysis}
\label{sec:method}
An obvious signature of the Earth obscuration of the CXB in the
INTEGRAL data is the characteristic depression in the observed light
curves (e.g. Fig.\ref{fig:aurora}). This depression to the first
approximation reflects variations of the angular size of the Earth as
seen by each of the instruments. Unfortunately there are several
other variable components contributing to the light curves, which
complicate the extraction of the CXB signal. We now discuss all these
components and argue that their contribution can be suppressed by
filtering the data or can be explicitly taken into account. 

\subsection{Model components}
The total flux $F(E,t)$ measured by the INTEGRAL detectors at a given
time $t$ and given energy $E$ can be separated into several
components: internal detector background $B(E)$, flux from bright
galactic X-ray sources $\sum_i F_{S,i}(E,t)$, collective flux from
weak unresolved galactic sources (Galactic Ridge) $F_{\rm
Ridge}(E,t)$, CXB emission $F_{\rm CXB}(E,t)$, Earth atmospheric
emission induced by cosmic ray particles $F_{\rm Atm}(E,t)$, the Earth
auroral and day side emission $F_{\rm Aur}(E,t)$, the emission (CXB
and galactic sources) reflected by the Earth atmosphere $F_{\rm
Refl}(E,t)$. The contributions of these components have to be
evaluated with account for the Earth modulation and the effective area
of the detectors. Each of the components leaves a specific signature
in the detector light curve.

Accordingly we can simply write that the flux is the sum of the above
mentioned components:
\begin{eqnarray}
F(E,t)=B(E)+\sum_i F_{S,i}(E,t)+F_{\rm Ridge}(E,t)+\nonumber\\ F_{\rm
Aur}(E,t)+F_{\rm Atm}(E,t)+F_{\rm CXB}(E,t)+F_{\rm Refl}(E,t).
\label{eqn:components}
\end{eqnarray}
Here and below we use the notion $F(E)$ for the count rates (in $\rm
cnts~s^{-1}~keV^{-1}$) measured by each of the detectors in a given
energy bin and $S(E)$ for true spectra (in $\rm
phot~s^{-1}~cm^{-2}~keV^{-1}$). The conversion from $S(E)$ to $F(E)$
is the convolution with the effective area, vignetting term, 
solid angle (for diffuse sources)  and the energy response
matrix of each detector. For brevity we do not explicitly write the
convolution in the subsequent expressions.

\subsection{CXB model}
The canonical broad band CXB spectrum in the energy range of
interest is based on the HEAO-1 data. The following analytic
approximation was suggested by Gruber et al. (1999):
\begin{eqnarray}
S^{G99}_{CXB}(E)=\left\{\begin{array}{ll}
7.877~E^{-0.29}~e^{-E/41.13} & 3<E<60~{\rm keV} \\
& \\
0.0259~(E/60)^{-5.5}+ & \\
0.504~(E/60)^{-1.58}+ & E>60~{\rm keV} \\
0.0288~(E/60)^{-1.05} &
\end{array}
\right. 
\label{eqn:cxb}
\end{eqnarray}
 Here $S^{G99}_{\rm CXB}(E)$ is in units of ${\rm keV/keV~cm^{-2}~s^{-1}~
sr^{-1}}$. Below we use notation $S_{\rm CXB}(E)$ for the (unknown)
CXB spectrum to distinguish it from  $S^{G99}_{\rm CXB}(E)$ spectrum
given by eq.\ref{eqn:cxb}.

The obscuration of the CXB by the Earth disk produces the depression
in the recorded flux with an amplitude set by the surface brightness
of the CXB and the solid angle $\Omega(t)$ subtended by the Earth
$S_{CXB}(E) \Omega(t)$.

\subsection{CXB reflection by the Earth's atmosphere}
\begin{figure}
\includegraphics[width=\columnwidth]{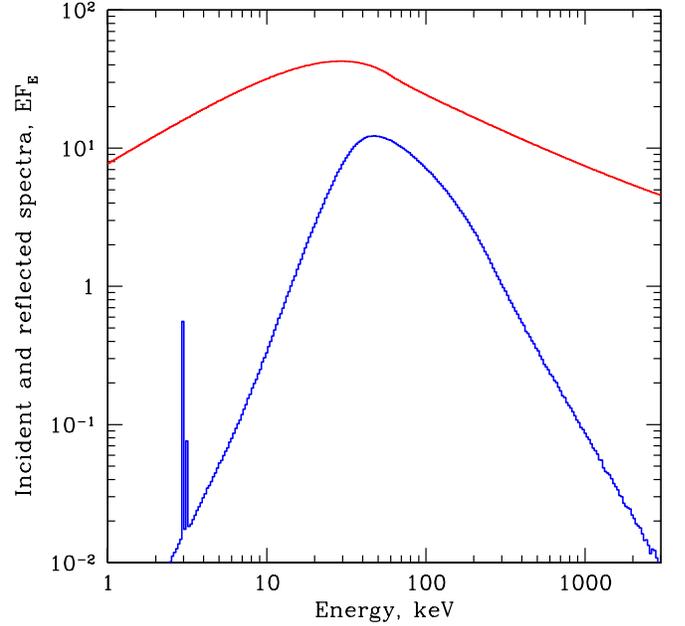}
\caption{The CXB spectrum and the spectrum reflected by the Earth
atmosphere (Churazov et al., 2006). The reflected spectrum was
integrated over all angles. The features in the reflected spectrum near
3 keV are the fluorescent lines of argon. The approximation of the albedo given in eq. \ref{eqn:refl} neglects the flourescent lines. 
\label{fig:albedo}
}
\end{figure}

The outer layers of the Earth's atmosphere reflect part of the
incident X-ray spectrum via Compton scattering. The picture is very
similar to the well studied case of the reflection from a star surface
(e.g. Basko, Sunyaev \& Titarchuk, 1974) or an accretion disk
(e.g. George \& Fabian, 1991) except for the different chemical
composition of the reflecting medium. The spectrum reflected by
the Earth's atmosphere was calculated (Churazov et al., 2006) using a
Monte-Carlo method for a realistic chemical composition of the
atmosphere and taking into account all relevant physical processes
(photoabsorption, Compton scattering and Rayleigh scattering on the
electrons bound in the molecules and atoms). The reflection is most
significant near $\sim$60 keV and declines towards lower or higher
energies (see Fig.\ref{fig:albedo}).

 The ratio of the reflected and incident spectra (energy dependent
albedo $A(E)$) was evaluated for the CXB spectrum shape measured by HEAO-1
(eq.\ref{eqn:cxb}) and approximated by the following expression (eq.6
of Churazov et al. 2006):
\begin{eqnarray}
A(E)=\frac{1.22}
{\left(\frac{E}{28.5}\right)^{-2.54}+\left(\frac{E}{51.3}\right)^{1.57}-0.37}\times
\nonumber \\
\frac{2.93+\left(\frac{E}{3.08}\right)^4}{1+\left(\frac{E}{3.08}\right)^4}\times
\nonumber \\
\frac{0.123+\left(\frac{E}{91.83}\right)^{3.44}}{1+\left(\frac{E}{91.83}\right)^{3.44}}
\label{eqn:refl}
\end{eqnarray}

For an observer located at some distance $D$ from the Earth the flux
integrated over the full Earth disk will then be:
\begin{eqnarray}
F_{\rm Albedo}(E)=S_{CXB}(E)A(E)\Omega(t),
\label{eqn:albedo1}
\end{eqnarray}
 Strictly speaking the factorization of the reflected spectrum
into the product of the incident spectrum and the energy dependent
albedo is valid only if $A(E)$ was calculated for a spectrum having the
same spectral shape as $S_{CXB}(E)$. In practice the albedo is not
strongly sensitive to the assumed shape of the incident spectrum (see
Churazov et al., 2006 for details). Much of our results (see below)
were obtained assuming that $S_{\rm CXB}=\beta~S^{G99}_{\rm CXB}(E)$
which makes the expression (\ref{eqn:albedo1}) exact. Here $\beta$ is
the normalization constant, independent of energy. The angular
dependence of the emerging spectra is not important as long as the
full Earth disk is used (see Churazov et al., 2006). In particular the
result does not depend on the distance $D$ from the Earth. If however
only a part of the disk is seen or if the effective area of the
telescope varies substantially across the Earth disk, then the albedo
is no longer described by a universal function and has to be
calculated taking into account the angular dependence of the emerging
radiation. For IBIS and SPI instruments with wide FoVs  this
simple approximation may fail only for the very beginning (few ksec)
or the very end of each observation when only fraction of the disk is
seen (Fig.\ref{fig:obsmode}). As is discussed further in section
\ref{sec:compact} the data during first 5 ksec of each observation
have been discarded anyway. At the end of each observation the effect
is small, since the Earth has much smaller angular size
and vignetting near the zero sensitivity FoV edge further suppresses
the signal. JEM-X has narrower FoV, but the contribution of the
reflection is small in the JEM-X energy range. Thus the modulation of
the CXB flux by the Earth (depression in the light curve) observed by
the instruments can be written as:
\begin{eqnarray}
F_{\rm CXB}(E,t)=S_{\rm CXB}(E)[1-A(E)]\Omega(t).
\label{eqn:albedo2}
\end{eqnarray}
Since the outer layers of the Earth atmosphere may be opaque at low
energies and transparent at high energies the apparent angular
``size'' of the Earth $\Omega$ (see eq.\ref{eqn:albedo2}) does depend on
energy. For instance at 1 keV the optical depth reaches unity for a line of
sight having an impact parameter of $\sim R_e+$ 120 km, where $R_e$ is the Earth radius. For comparison at 1 MeV the
corresponding impact parameter is $\sim R_e+$ 70 km. This effect
limits the accuracy of the above approximation - eq.\ref{eqn:albedo2}
- to $\sim$1-2\%. 

\subsection{Earth X-ray albedo due to Galactic sources}
The Earth atmosphere will also reflect X-rays from the Galactic
sources. Since the pointing direction is in the general direction of
the Galactic Center, most of the Galactic bright sources are located
``behind'' the Earth and their reflected emission does not contribute
to the flux measured by INTEGRAL. One can therefore expect that the
Crab nebula, located at the Galaxy anti-center will be the dominant
source of the contamination. An easy estimate of this contamination
can be done simply by calculating the total incident flux of the CXB
and Crab emission coming on to a side of the Earth towards the
observer. If we assume that the intensity of the CXB at
the energy of $\sim$ 30 keV is $\sim 1.8$ mCrab per square degree, and
we integrate over all incidence angles, we obtain an estimate of the
total incident CXB flux per 1 ${\rm cm^2}$ of the atmosphere of $\sim$ 19 Crab.
 Thus for one
hemisphere the ratio of total incident CXB and Crab fluxes is
$\frac{19\times 2\pi R_e^2}{\pi R_e^2}\sim 40:1$. The maximal value
of the albedo is of the order of 35\%. Therefore the contribution of
the Crab emission reflected by the Earth atmosphere can be roughly
estimated to be at the level of $35\%/40\sim$1\% relative to the CXB
signal. In the subsequent analysis we neglect this component.

\subsection{Earth atmospheric emission}
\label{sect:atm}
Due to the bombardment by cosmic rays, the Earth's atmosphere is a
powerful source of hard X-ray and gamma-ray emission. Although
experimental and theoretical studies of this phenomenon have a long
history starting in the 1960s, there remains a significant uncertainty
with regard to the emergent spectrum of the atmospheric emission, in
particular in the energy range of interest to us -- between 10 and
200~keV. On the other hand, the hadronic and electromagnetic processes
responsible for the production of atmospheric emission, although
complicated, are well understood. Similarly, the incident spectrum of
cosmic rays has been recently measured with high precision. This
implies that with the power of modern computers, the spectrum and flux
of atmospheric X-ray emission should be predictable by simulations to
a reasonable accuracy. We performed such a numerical modeling using
the toolkit Geant 4. A detailed report on our analysis and results is
presented elsewhere (Sazonov et al. 2006). Here we briefly summarize
the main outcome of our simulations.

We found that at the energies of interest to us, most of the emerging
X-ray photons have almost no memory of their origin, i.e. they were
produced (mainly by bremsstrahlung and positron--electron
annihilation) with relatively high energy at a significant atmospheric
depth (several grams or tens of grams of air per sq. cm from the top of the
atmosphere) and escaped into free space after multiple Compton
down-scatterings. 

This process is similar to the formation of supernova hard X-ray
spectrum resulting from down Comptonization of $^{56}$Co gamma ray
lines in an optically thick envelope.  Such a spectrum was observed
from SN1987A \citep[][]{sunyaev87}

\begin{figure}
\includegraphics[width=\columnwidth]{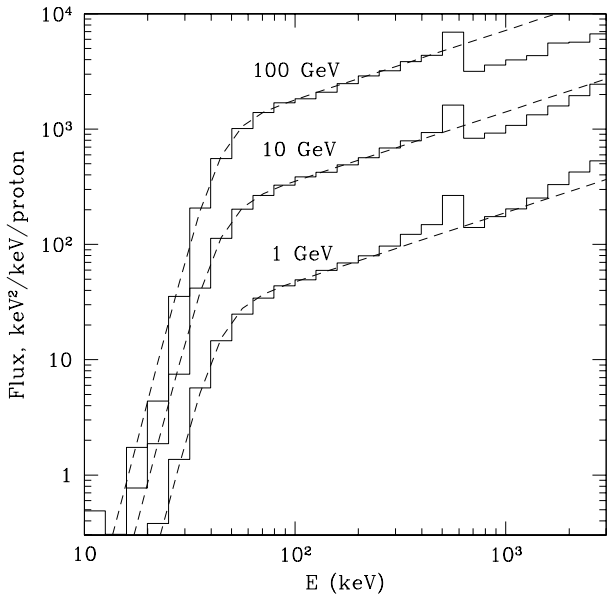}
\caption{Examples of simulated spectra (solid lines) of atmospheric
emission produced by cosmic protons of given energy: $E_p=$ 1, 10 and
100 GeV. It can be seen that in the photon energy range 25-300 keV the
shape of the emergent spectrum is almost invariant and well fitted with
eq.\ref{eq:atm_fit} (the dashed lines).
\label{fig:atm}
}
\end{figure}

As a result, the emergent spectrum (see Fig.\ref{fig:atm}) is barely
sensitive to the energy of the parent cosmic ray particle (in the
relevant range from a fraction of 1~GeV to a few hundred GeV) or to
the type of the incident particles (e.g. proton, alpha-particle,
electron). In the energy range 25--300~keV the emergent spectrum is
well fitted with the following formula (Sazonov et al., 2006):

\begin{equation}
S_{ATM}(E)=\frac{C}{(E/44)^{-5}+(E/44)^{1.4}}
\label{eq:atm_fit}
\end{equation}

The photon spectrum peaks around 44~keV. At
energies below the maximum the spectrum shows a rapid decline
(as $E^{-5}$) due to photoabsorption. Since the emission is already very
low at 25~keV, we did not try to approximate the spectrum below this
energy (where the shape is noticeably different). Above $\sim 300$~keV
the spectrum starts to be dominated by the comptonization continuum
associated with the prominent 511~keV annihilation line and the
approximation given by equation~(\ref{eq:atm_fit}) again breaks down.

The work of Sazonov et al. (2006) also predicts the normalization $C$
of the atmospheric emission spectrum as function of the solar
modulation parameter and the rigidity cut-off associated with the
Earth's magnetic field. The cut-off is calculated in the shifted
dipole approximation, with Stoermer's formula. For the specific
conditions of the INTEGRAL observations, the predicted brightness of
the atmospheric emission is 31.7 ${\rm
keV^2~cm^{-2}~s^{-1}~keV^{-1}~ sr^{-1}}$. This value was obtained by
averaging the brightness of the atmosphere over the Earth disk and
adopting the solar modulation parameter $\phi=$0.45 GeV, corresponding
to the dates of the observations (see Sazonov et al. 2006 for
details). This value will be compared with that inferred from our
fitting procedure of the measured spectrum in Section
\ref{res:spectra}.

\subsection{Earth Auroral and day-side emission}
\label{sec:aurora}
As mentioned above not all JEM-X observations can be well described by
a simple model of the CXB signal modulated by the Earth. In the
revolutions 404 (see Fig.\ref{fig:aurora}, bottom panel) and 405 the
light curves clearly deviate from the prediction of the pure CXB
modulation model. The excess in the light curve appeared when a large
part of the field of view was covered by the Earth disk. This suggests
that there is an additional source of the X-ray emission in the
direction of the Earth. The most plausible explanation of this excess
is the Earth auroral emission. We then added to our
CXB-modulation model an 
additional component, assuming that the emission comes from the
circular region with a radius of order of $0.1\times R_{E}$ around the
Earth North magnetic pole. This two-component model 
provides a reasonable description of the data (see
Fig.\ref{fig:aurora}), thus broadly supporting the auroral origin of
the emission. However the observed light curve in revolution 404
(Fig.\ref{fig:aurora}, bottom panel) suggests that a more
complicated model of auroral emission is needed.  We therefore decided
to limit the analysis of the JEM-X data to revolutions 401 and 406 which
are not affected by the auroral emission.

During our observations the Earth disk was predominantly dark (see
Fig.\ref{fig:obsmode}). The emission from the dark side (induced by
the cosmic rays bombardment) is discussed in section
\ref{sect:atm}). The day side of the Earth is a source of additional
X-ray emission due to the reflection of Solar flares and non-flaring
corona \citep[e.g.][]{itoh02}. This emission is typically very soft
and is not important at energies higher than a few keV. In principle,
an increased hard X-ray flux from dayside Earth might be expected at the
time of powerful Solar flares, but monitoring of the Solar activity
(http://www.sec.noaa.gov) did not show any significant Solar
flares during our observations.  In the subsequent analysis we have
neglected the Earth day side emission.

\subsection{Compact sources and the Galactic ridge}
\label{sec:compact}
As was already mentioned above both relatively bright compact sources,
which are detectable with INTEGRAL and unresolved Galactic emission
(Galactic Ridge) have to be removed from the data in order to get
a clean estimate of the CXB flux.

Bright compact sources can be detected directly by INTEGRAL
telescopes, and their contribution can be subtracted from the
detector's count rate. However this procedure would increase the
statistical errors of the CXB flux measurements, especially if a
large sample of compact sources is considered.

Contribution of the unresolved Galactic background (Galactic ridge
emission) can be estimated using the results of Revnivtsev et
al. (2006) and Krivonos et al. (2007). There it is shown that the
Galactic ridge X-ray emission surface brightness correlates very well with
the Galactic near-infrared surface brightness. 
Using data of COBE/DIRBE (LAMBDA archive , http://lambda.gsfc.nasa.gov/)
and correlation coefficients from
Krivonos et al. (2007) we obtained that the peak surface
brightness of the Galactic ridge at Galactic longitude $l\sim330$ is
approximately $\sim1$ mCrab/deg$^2$ in the energy band 17-60 keV,
providing approximately $\sim20-30$ mCrab net flux for IBIS/ISGRI and
SPI. Thus the contribution of the Galactic Ridge
can be modeled using the near-infrared data, convolved with the
angular response of the INTEGRAL instruments. 

\begin{figure}
\includegraphics[width=\columnwidth]{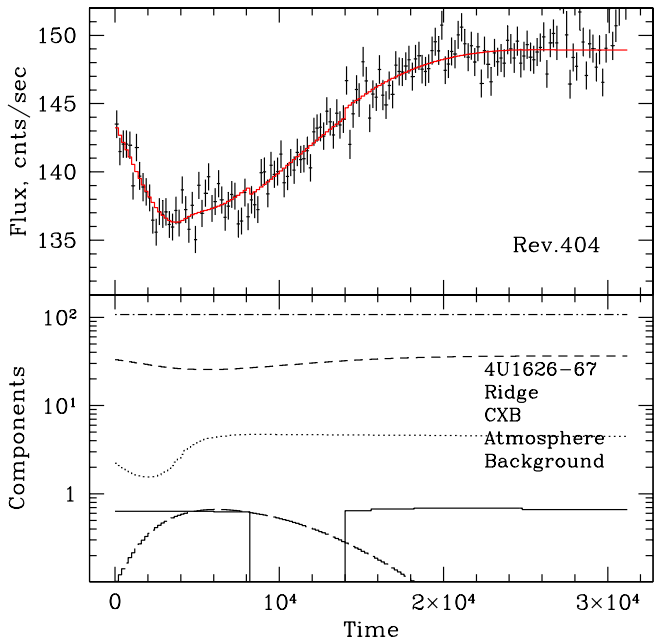}
\caption{Illustration of the various components contributing the the IBIS/ISGRI
light curve in the 20-40 keV band. Upper panel: total count rate in
the 20-40 keV band together with the best-fit model (solid curve).  Lower
panel: model components: internal detector background (dash-dotted
line), CXB emission (short-dash line), Galactic Ridge emission (dotted
line), the Earth atmospheric emission (long-dash line) and a single
compact source 4U1626-67 (solid line). 
The normalizations of these
components were free parameters of the model. The CXB flux, modulated by the
Earth disk, is the main source of the flux variations observed by 
INTEGRAL instruments. The internal detector background is assumed to
be stable.  The Galactic Ridge
crosses the edge of the IBIS field of view and the Earth blocks the
ridge emission only during first 3-4 ksec of the observation (see
Fig.1). Both CXB obscuration and the Earth atmospheric emission reach
the maximum amplitude approximately 6 ksec after the start of the
observation, when the Earth disk has still a large angular size and 
fills the central, most sensitive, part of the FoV.
\label{fig:components}
}
\end{figure}

\begin{figure}
\includegraphics[width=\columnwidth]{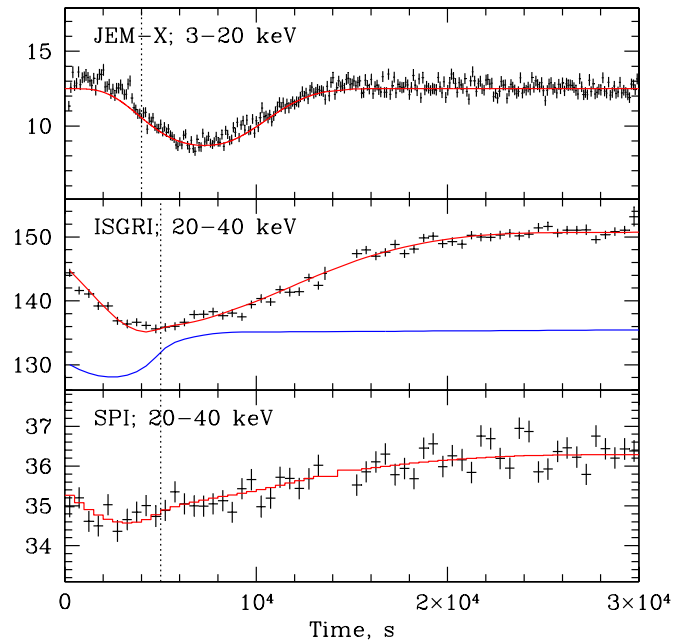}
\caption{
\label{fig:lcurves}
The light curves of JEM-X, IBIS/ISGRI and SPI instruments in units of
counts per second. In the middle panel the blue curve shows
schematically (with arbitrary normalization) the time dependence of
the Galactic Ridge emission, modulated by the Earth occultation. In
order to avoid contamination of the CXB measurements due to Galactic
plane contribution the first few ksec of data (on the left from the
vertical lines) were dropped from the analysis. Note that for JEM-X
less strict cut was applied, since its field of view is smaller than
that for the other instruments (see section \ref{sec:jem}).}
\end{figure}

A typical time dependence of the various components contributing to the
A typical time dependence of the various components contributing to the
total count rate of IBIS/ISGRI is illustrated in
Fig.\ref{fig:components}. In the upper panel the total count rate in the
20-40 keV band is shown together with the best-fit model. The 5 model
components, used in this illustration, are shown in the lower panel:
internal detector background, CXB emission, Galactic Ridge emission,
the Earth atmospheric emission and a single compact source
4U1626-67. The normalizations of each of these components were free
parameters of the model. Notice that since the surface brightness of the
Galactic Ridge drops sharply towards higher latitudes, this component
is affecting only the first few ksec of data. As one can see from
Fig.\ref{fig:obsmode} and \ref{fig:image_isgri} many compact sources
are clustered near the Galactic Plane, sharing the same area as the
Galactic Ridge. Adding the contributions of the Ridge and several
compact sources as independent components would make the task of
separating them very difficult. In order to make the analysis more
robust we decided to cut out the first $\sim$ 5 ksec of data (when the
Ridge contribution is not negligible) from further analysis as shown
in Fig.\ref{fig:lcurves}. This also removes much of problems with the
occultation of compact sources, located near the Plane. The only
moderately strong source located far above the Galactic plane is
4U1626-67. Inclusion/exclusion of this source in the model changes the
net 20-40 keV CXB flux by $\sim 2$\%. Therefore with the above data
selection all compact sources can be dropped from the model without
much impact on the final CXB spectra. Note that JEM-X has smaller
field of view, which does not cover the Galactic plane. The brightest
source in JEM-X field of view -- 1H1556-605 has a mean flux 5-7 mCrab
in JEM-X energy band and it is very far from the center of the
instrument field of view ($\sim5.6^\circ$) where the effective area of
the instrument drops to less than 5\% of the maximum. The estimated count
rate which might be caused by this source is below the Poisson noise of
the detector.  Therefore the data cut used for JEM-X is mainly
driven by the stability of the internal detector characteristics (see
section \ref{sec:jem}).

\subsection{Instruments absolute and cross- calibration}
\label{sect:crab}
CXB absolute flux measurements were done several times over the past
three decades \cite[see e.g.][]{horstman74,mazets75,kinzer78,
marshall80,mccammon83,kinzer97,vecchi99,
gruber99,kushino02,lumb02,revnivtsev03,revnivtsev05,hickox05}. The
reported fluxes show considerable scatter, part of which is likely
related to the problem of the absolute flux calibration of the
instruments. The same problem appears when the fluxes of the Crab
nebula (assumed to be a ``standard candle'') derived by different
instruments are compared
\citep[e.g][]{toor74,seward92,kuiper01,kirsch05}. Therefore in order
to make a fair comparison of the CXB spectrum obtained by different
instruments one has to make sure that the same assumptions on the
``standard candle'' (Crab) spectrum are made.

\begin{figure}
\includegraphics[width=\columnwidth]{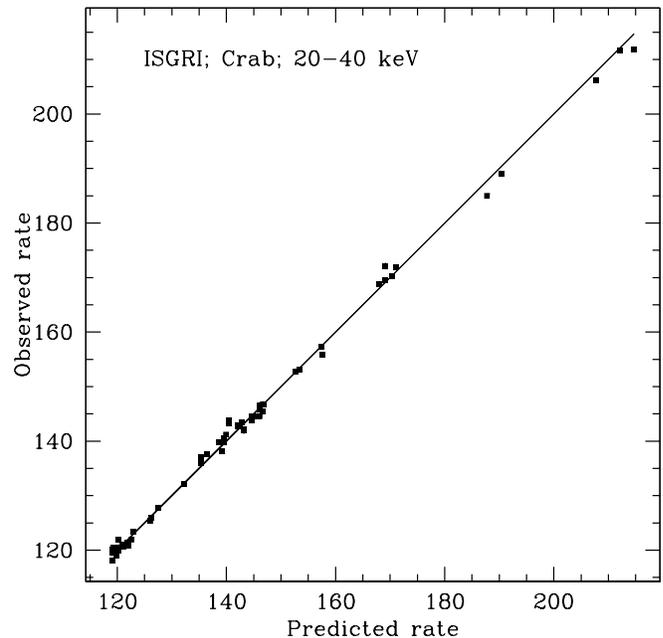}
\caption{Comparison of the predicted and observed 20-40 keV fluxes in
individual (few ksec long) observations during revolution 170. The
model included two components: constant in time detector background
and the predicted signal based on the source position in the IBIS/ISGRI
field of view.
\label{fig:crab_isgri2}
}
\end{figure}

We analyzed several Crab observations with INTEGRAL exactly in the
same way as we did for the Earth observations. In particular the
revolutions 170 and 300 were used. During these revolutions a large
number of observations was performed with the source (Crab) position
varying from almost on-axis to essentially outside the field of
view. The light curves in narrow energy bands were constructed,
together with the predicted signal (based on the source position and
the IBIS/ISGRI and SPI angular responses). The intrinsic detector
background was added as an independent component to the model (single
component for IBIS/ISGRI and 17 independent components for the 17 SPI
detectors). The internal background was assumed to be constant with
time. The Crab spectrum in counts/sec was derived from the best-fit
normalization of the model in each energy bin. For illustration the
comparison of the predicted and observed 20-40 keV fluxes in
individual (few ksec long) observations during revolution 170 is shown
in Fig.\ref{fig:crab_isgri2}.  Comparison of the spectra obtained in
revolutions 170 and 300 shows very good agreement between the
resulting spectra. Since the set of Crab positions in these two
revolutions were different we concluded that our model is providing
robust resulting source spectra when a large number of quasi-random
source positions is used. This kind of analysis is very similar to the
analysis of CXB occultation by the disk of the Earth and it is
therefore possible to use the derived raw Crab spectra for
calibration/cross-calibration purposes.

\begin{figure}[htb]
\includegraphics[width=\columnwidth,bb=35 60 553 400,clip]{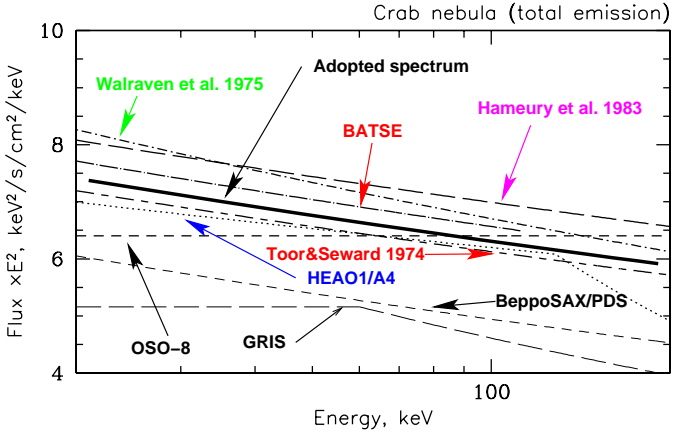}
\caption{Best-fit spectra of the Crab nebula (total emission) measured by 
different experiments in the hard X-ray band. The spectrum adopted
here is shown with the thick solid line. 
\label{fig:crabs}
}
\end{figure}

 Most of the historic measurements of the Crab spectrum
(the pulsar plus nebulae) suggest that a single power law is a
reasonable approximation below $\sim$100 keV
\citep[e.g][]{toor74,seward92,kuiper01,kirsch05} with the scatter in
the reported values of the photon index of $\sim$0.1 (see Fig.\ref{fig:crabs}).  The reported
fluxes also show considerable scatter (Table \ref{tab:crabs}).  While
the accurate absolute measurements of the Crab spectrum are of
profound importance for the X-ray astronomy, for our immediate
purposes the crucial issue is to use the same definition of the
``standard candle'' for all instruments to allow for a fair comparison
of the CXB signal. The basic assumption here is that the
intrinsic variability of the Crab is small compared to the required
level of accuracy. 

We choose the Crab spectrum in the form $dN/dE=10~E^{-2.1}$
phot~s$^{-1}$~cm$^{-2}$~keV$^{-1}$. This simple approximation provides
a reasonable compromise among historic Crab observations in terms of
the spectral slope and flux in the energy range of interest (see
Fig.\ref{fig:crabs} and Table \ref{tab:crabs}). We then cross-calibrated the
INTEGRAL instruments (using the spectra extracted with the above
mentioned procedure) to ensure that the reference Crab spectrum is
recovered. Since the slope of CXB is not dramatically different from
the Crab spectrum near the energy of the maximum energy release, one
can do this procedure by introducing a special fudge factor (a smooth
function of energy) such that the raw spectra in counts divided by the
effective area with this fudge factor recover the assumed Crab
spectrum in phot~s$^{-1}$~cm$^{-2}$~keV$^{-1}$.  While this approach
has obvious limitations and disadvantages its simplicity and
robustness fits well the purposes of this particular work (at least in
the first approximation).

Thus, in our subsequent analysis we assume that the shape of the Crab
nebula spectrum in the energy band 5-100 keV is described by a power
law $dN/dE=10~E^{-2.1}$ phot~s$^{-1}$~cm$^{-2}$~keV$^{-1}$. Any
changes in the assumed Crab normalization and in the spectral shape
can then be easily translated into changes in the CXB spectrum.  
While reasonable, this choice of the Crab spectrum is still arbitrary
and one has to bear this in mind when making the comparison with the
results of other mission. An assumption that any of the values quoted
in Table \ref{tab:crabs} have equal likelihood of being the closest
approximation to the true (unknown) flux from the Crab implies the
uncertainty in absolute flux calibration can be as high as $\sim$30\%
(e.g. flux measured by GRIS), although majority of points come within
$\pm$10\% of the value adopted here. A more fair relative comparison
is possible only if the measured CXB flux is quoted together with the
measured Crab flux in the same energy band.

\begin{table}
\caption{Crab nebula fluxes in the 20-50 keV energy band, measured by different
experiments. The value adopted for this paper is shown in bold face.}
\begin{tabular}{l|c}
Measurements& Flux 20-50 keV\\
 &$10^{-9}$erg s$^{-1}$ cm$^{-2}$\\ 
\hline
Rockets\citep{toor74}&10.0\\
Rocket\citep{walraven75}&11.2\\
OSO-8 \citep{dolan77}&9.4\\
\cite{hameury83}&11.3\\
HEAO1/A4\citep{jung89}& 9.8\\
GRIS\citep{bartlett94}&7.6\\
CGRO/BATSE\citep{ling03}&10.8\\
BeppoSAX/PDS\citep{kirsch05}&8.6\\
RXTE/HEXTE\citep{kirsch05}&10.6\\
{\bf Adopted here} $dN/dE=10~E^{-2.1}$&10.4\\
\hline
\end{tabular}
\label{tab:crabs}
\end{table}


\subsection{CXB cosmic variance}
The total CXB flux due to unresolved sources within the given area of
sky may vary due to Poissonian variations of the number of sources,
intrinsic variability of the source fluxes or due to the presence of a
large scale structure of the Universe \cite[see e.g.][]{fabian92}.
The number counts of extragalactic sources at about the flux level
corresponding to the INTEGRAL instruments sensitivity ($f_{\rm x} \sim
10^{-11}$ erg~s$^{-1}$cm$^{-2}$ in the 20-50 keV band) is consistent with a power law having a slope
$\alpha=-1.5$ and normalization $\sim1.4\times10^{-2}$ deg$^{-2}$
\citep{krivonos05}. The effective size of the region of sky occulted
by the Earth during INTEGRAL observations is $\sim 10\times 15$
degrees (visible diameter of the Earth times the length of the path
the Earth center makes in the sensitive part of the FOV). In this case
expected Poissonian variations of the CXB flux due to unresolved
extragalactic sources are smaller than 1\%. This estimate assumes that
sources with a flux $>10^{-11}$ erg~s$^{-1}$~cm$^{-2}$ would be detected and
accounted for. A similar estimate can be obtained by rescaling the
RXTE/PCA 3-20 keV measurement of the CXB variations at 1$^\circ$
angular scale
\citep{revnivtsev03} to a larger area.

The variance of the CXB originating from large scale structure of
the local Universe was extensively studied using the HEAO-1 data
\citep[e.g.][]{shafer83,miyaji94,lahav97, treyer98,scharf00}. The
largest scale anisotropy (dipole component) is approximately
consistent with Compton-Getting effect due to the motion of the Earth
with respect to the Cosmic Microwave Background rest frame. However
some additional dipole anisotropy was suggested, which is consistent
with the predicted large-scale structure variations
\cite[e.g][]{treyer98,scharf00}. It was shown that in total these
variations are at the level of $\sim0.5$\%. At smaller angular scales
the amplitude of the variations due to large scale structure is rising
but at the angular scales $\sim10^\circ$ it should not exceed
$\sim$1\% \citep[e.g.][]{treyer98}.  Summarizing all of the above one
can conclude that for a region of sky used for a determination of the
CXB spectrum during the INTEGRAL observations (effective size of the order
of $10\times 10$ degrees), the CXB cosmic variance is becoming a
limiting factor at a level of accuracy of $\sim$1\%.

\section{Results}
\label{res:spectra}
\begin{figure*} 
\includegraphics[width=\columnwidth]{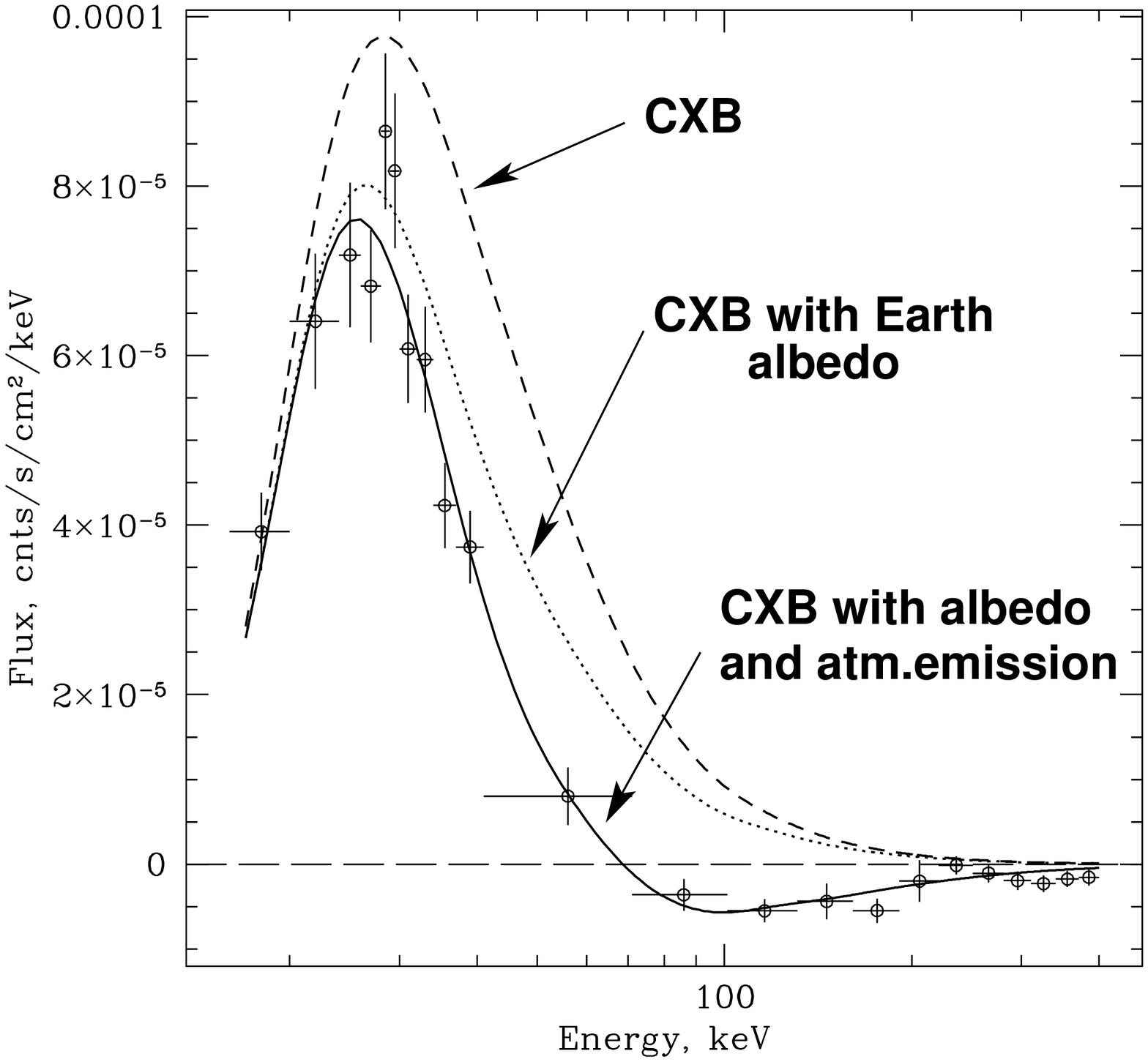}
\includegraphics[width=\columnwidth]{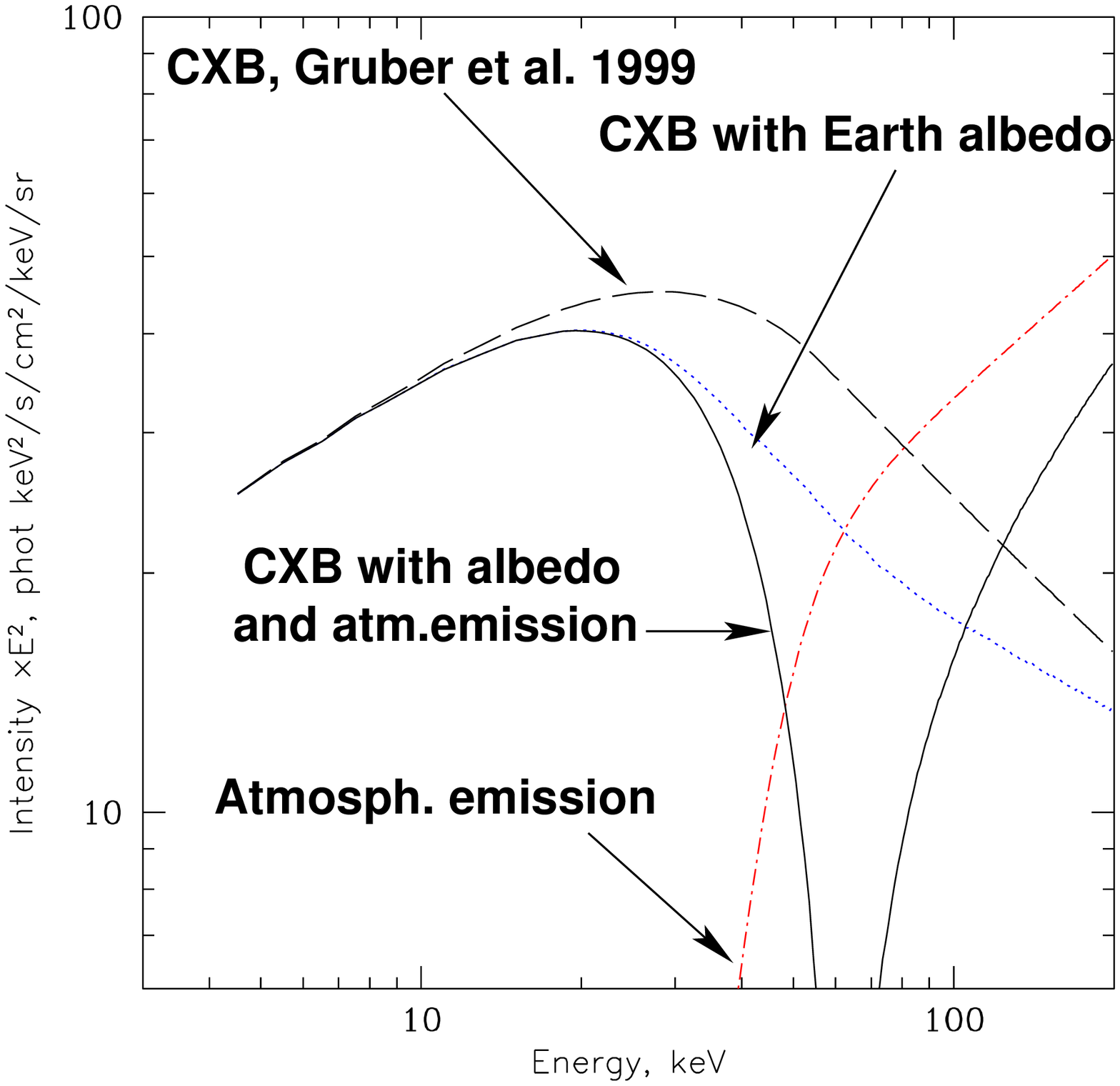}
\caption{Left -- the $F_{\rm Earth}(E)$ spectrum obtained by
INTEGRAL/SPI. The dashed line shows the CXB model spectrum. The dotted
line shows the same spectrum modified by the Earth albedo. The solid line
show the expected spectrum corrected also for atmospheric emission.
Right -- theoretical models of the same spectral components, used in the
left figure: CXB spectrum, CXB spectrum corrected for albedo. In
addition the atmospheric emission emission spectrum is also shown (the
dashed-dotted line). Note that the CXB and the
atmospheric emission components contribute to the observed spectrum
with different signs and the total signal (CXB with albedo and
atmospheric emission) changes sign between 60 and 70 keV.
\label{fig:demo}
}
\end{figure*}

As discussed in the previous section after the data filtering only 4
components are left in the model: the intrinsic detector background,
the modulation of the CXB signal, the CXB radiation reflected by the
Earth and the Earth atmospheric emission. All these components
primarily depend on the solid angle (within the FOV) subtended by the
Earth. To evaluate the impact of each of these components on the
observed light curves one has to average the signals over the Earth
disk, taking into account the effective area of the
telescopes. E.g. for the atmospheric emission the signal was averaged
taking into account rigidity distribution, angular distribution of the
emerging atmospheric emission and the telescopes vignetting. After
such averaging the CXB flux modulation (including Compton reflection)
and the flux variations due to the Earth atmospheric emission produce
very similar signatures (but with the opposite signs) in the detector
light curves.  This severely complicates any attempts to disentangle
these contributions directly from the observed light curves. Instead
it was assumed that a combination of these components can be described
as a single time dependent signal for each measured energy bin:
\begin{eqnarray}
F(E,t) \approx C(E)-S_{\rm
CXB}(E)\left[1-A(E)\right]\Omega(t)+\nonumber \\S_{\rm ATM}(E)\Omega(t) = C(E)-S_{\rm Earth}(E)\Omega(t),
\label{eqn:esum}
\end{eqnarray}
where $S_{\rm ATM}(E)$ is the spectrum of the cosmic ray induced
atmospheric emission (averaged over the Earth disk and normalized per
unit solid angle), and $S_{\rm Earth}(E)=S_{\rm
CXB}(E)\left[1-A(E)\right]-S_{\rm ATM}(E)$ is the combined flux of all
components related to the presence of the Earth in the FoV. A constant
in time $C(E)$ includes the intrinsic background and total combined
flux of Galactic sources and CXB without the Earth in the field of
view. Using the observed light curves $F(E,t)$ in each energy bin and
the known effective solid angle $\Omega(t)$ subtended by the Earth in
the FoV of each instrument the $F_{\rm Earth}(E)$ counts spectra
(i.e. $S_{\rm Earth}(E)$ convolved with the effective area and the
energy redistribution matrix) were obtained. As an example the counts
spectrum $F_{\rm Earth}(E)$ for SPI is shown in Fig.\ref{fig:demo}. 
In the right panel of Fig.\ref{fig:demo} the components of the
model spectrum $S_{\rm Earth}(E)$ (CXB spectrum, CXB spectrum
corrected for the Compton reflection and for the Earth atmospheric
emission) are shown. 

\begin{figure}
\includegraphics[width=\columnwidth,bb=10 40 550 500,clip]{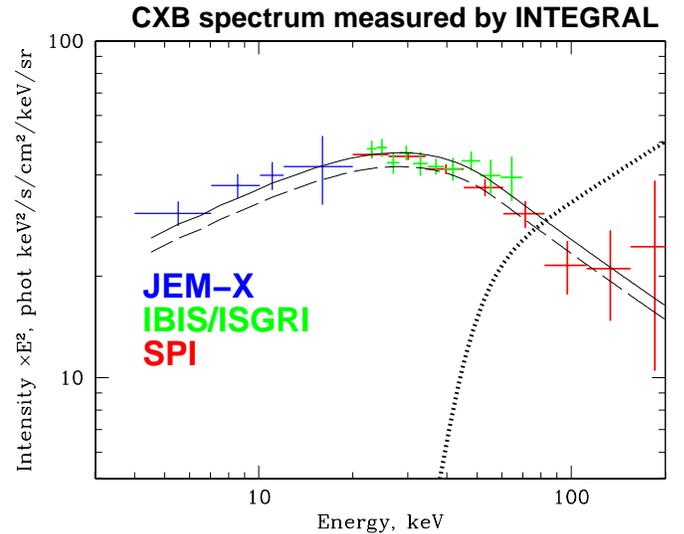}
\caption{Spectrum of the CXB measured by INTEGRAL instruments.  The
error bars plotted account for the uncertainties in the normalization
of the atmospheric emission component. The dashed line shows the
analytic approximation of the CXB spectrum by Gruber et al., 1999. The
solid line shows the same spectrum with the best-fit normalization
obtained in this work. The thick dotted line shows the best-fit
spectrum of the Earth atmospheric emission.
\label{fig:final}
}
\end{figure}

The observed spectrum was fitted in XSPEC (e.g.\citealt{xspec}) with a 3-component model, which includes: i)
the CXB spectrum in the form of eq.(\ref{eqn:cxb}) with free
normalization, ii) a fixed multiplicative model describing reflection
of the CXB from the Earth atmosphere according to 
eq.(\ref{eqn:refl},\ref{eqn:albedo2}), and iii) the Earth atmospheric
emission in the form of eq.(\ref{eq:atm_fit}) with free
normalization. The best-fit provides us two parameters of the model:
the normalization of the CXB spectrum and the normalization of the
Earth atmospheric emission. The CXB spectrum as measured with INTEGRAL
is shown in Fig.\ref{fig:final}. The data points shown in this figure were
obtained from the observed spectra in counts/s by subtracting the
best-fit atmospheric emission component, correcting for the Compton
reflection and for the effective area. For comparison the reference CXB
model spectrum is shown with the dashed line, with the absolute
normalization increased by 10\% compared to eq.\ref{eqn:cxb}. 
 There is a reasonable agreement between the data and the renormalized
CXB spectrum. More sophisticated models are not required by
the present data set. Below $\sim$10-20 keV extended data sets, 
available from other observatories (e.g. RXTE), provide
better constraints on the CXB shape than INTEGRAL. More relevant for
INTEGRAL observations is the consistency of the CXB spectrum
approximation on the energies above 30 keV, i.e. above the energy of
the CXB peak luminosity. Considering only the data in the 40-200 keV
range and approximating the spectrum with a power law one gets the
photon index of 2.42$\pm$0.4. For comparison, the CXB spectrum
measured by HEAO-1 (Gruber et al. 1999) can be characterized by a
photon index $\sim$2.65 in the same range.  If errors scale roughly as
square root of time then $\sim$4 times longer data set will be
required to bring the uncertainty in the photon index to
$\sim0.2$. Note that these estimates are based on the assumption that
the shape of the atmospheric emission is accurately predicted by
simulations. Further observations during different phase of the Solar
cycle would be very instrumental in proving this assumption.

Using the accurate measurement of the cosmic ray spectra and detailed
simulations (Sazonov et al., 2006) one can make an accurate prediction
of the atmospheric emission. In the present analysis we treat the
normalization of this component as a free parameter of the model.  The
best-fit normalization of this component obtained in present analysis
is $32.9 \pm 1.3$ ${\rm keV^2 cm^{-2} s^{-1} keV^{-1} sr^{-1}}$ which
agrees well with the expected value of $31.7$.  This excellent
agreement is encouraging.  We note however that the normalization of
this component is subject to the same systematic uncertainties
associated with the absolute flux calibration discussed above.  Further
observations with INTEGRAL (with a different seasonal modulation of
the cosmic ray spectrum) would be useful to verify the agreement of
observations and predictions.  Potentially the atmospheric emission
could become a useful absolute calibrator for the instruments
operating in the hard X-ray/gamma-ray bands.

\begin{figure}
\includegraphics[width=\columnwidth,bb=10 40 550 500,clip]{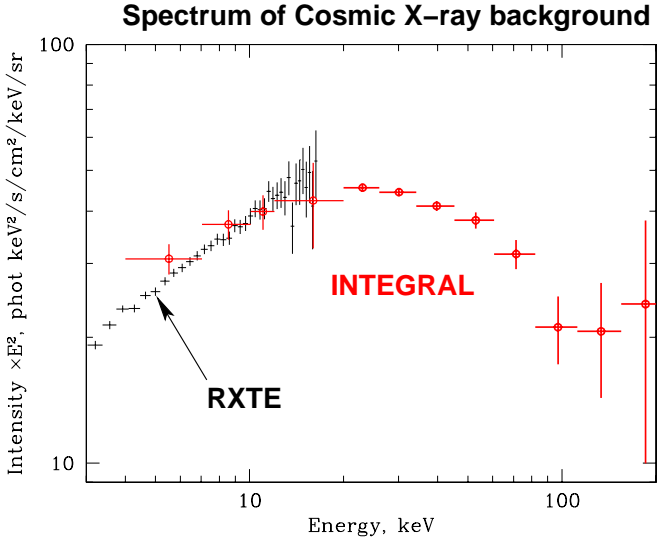}
\caption{Comparison of the CXB spectrum derived by INTEGRAL with the
3-20 keV CXB spectrum derived from the RXTE data
\citep{revnivtsev03}. The IBIS/ISGRI and SPI data points were averaged
in this plot.
\label{fig:comparison}
}
\end{figure}

In order to test the sensitivity of the results to the assumed
shape of the Crab spectrum, we repeated the same analysis varying the
assumed Crab photon index by $\pm 0.1$ (keeping the 20-50 keV flux
unchanged) and making appropriate changes in the efficiency fudge
factors. The resulting best-fit normalization of the CXB component
changed by $\pm\sim$1.2\%. This is of course an expected result, given
that the assumed 20-50 keV flux from the Crab nebula was unchanged.

 We now summarize the uncertainties in the derived CXB
normalization.  A pure statistical error (joint fit to JEM-X,
IBIS/ISGRI and SPI data) in the normalization of the CXB component is
$\sim$1\%. Further uncertainties are: neglected contributions from
compact sources $\le $2\%, modeling the atmospheric reflection $\sim
$1-2\%, uncertainty in the Crab photon index $\sim$1\%. On top of
these uncertainties, which if added quadratically 
amount to $\sim $3\%, comes the absolute flux calibration. From the
comparison of the Crab 20-50 keV flux measurements (Table
\ref{tab:crabs}) it is clear that this is by far the largest source of
uncertainty. In particular, as is mentioned above the flux measured by
INTEGRAL at $\sim$30 keV is $\sim$10\% higher than the value predicted
by the fit of Gruber et al., 1999. If we adopt the 20-50 keV Crab flux
measured by HEAO-1/A4 (Jung, 1989, see Table \ref{tab:crabs}) and
rescale the INTEGRAL measurement accordingly, this difference will
shrink to $\sim 4$\%.

In Fig.\ref{fig:comparison} we plot the INTEGRAL data together with
the CXB spectrum in the 3-20 keV derived from the RXTE data
\citep{revnivtsev03}. The RXTE data points are above the fit by Gruber
et al.(1999), and are derived using a Crab spectrum as
$dN/dE=10.8~E^{-2.1}$ phot~s$^{-1}$~cm$^{-2}$~keV$^{-1}$ in the 2-10
keV band, i.e. 8\% higher than is adopted in this paper. Recent
re-analysis of the HEAO-1 A2 data (Jahoda et al., 2007) gives the 2-10
CXB flux $\sim$ 10\% lower than the RXTE data points shown in
Fig.12. At the same time the most recent Chandra measurements (Hickox
\& Markevitch, 2006) yields the 2-8 CXB flux higher (but consistent
with 1$\sigma$) than the RXTE flux of Revnivtsev et al., 2005. Clearly
at present the absolute flux calibration of the instruments (both in
the standard 2-10 and hard 20-100 keV X-ray bands) is the dominate
source of uncertainties/discrepancies in the CXB measurements.

\section{Conclusions}
Using the modulation of the aperture flux by the Earth disk, the
INTEGRAL observatory measured the spectrum of the cosmic the X-ray
background in the energy range $\sim$5-100 keV.  The observed flux
near the peak of the CXB spectrum (in the $\nu F_\nu$ units) at 29 keV
is 47 ${\rm keV^2 cm^{-2} s^{-1} keV^{-1} sr^{-1}} \pm 0.5 ~(\pm 1.5
)$. The qouted uncertainties are pure statistic errors and (in
parentheses) systematic errors, excluding the uncertainty associated with
the choice of the model Crab spectrum. This value is $\sim$10\%
higher than suggested by Gruber et al., 1999. Note that in the present
analysis the absolute flux calibration was done assuming that the
spectrum of the Crab nebula in the relevant energy band is described
by the expression $dN/dE=10~E^{-2.1}$
phot~s$^{-1}$~cm$^{-2}$~keV$^{-1}$. Any changes in the CXB
normalization directly translates into the changes in the energy
release by the supermassive black holes in the Universe. These numbers
are important for estimating the radiative efficiency of the growing
black holes.

In the present analysis the observed level of the atmospheric emission
was found to be very close (within 10\%) to the results of the
simulations. Since accurate measurements of the cosmic ray spectra are
now available the Earth atmospheric emission could become a useful
``calibration'' source for the instruments operating in the few
hundred keV range.

The present observations were made during Solar minimum when the
expected level of the atmospheric emission (due to cosmic rays
interactions with the atmosphere) is close to the maximum. The
``background'' field was close to the Galactic Plane and part of the
data was discarded to avoid contamination of the signal by Earth
obscuration of the unresolved Galactic sources.  Future (longer)
observations during Solar maximum and with the pointing direction away
from the Galactic Plane would be very useful to verify the robustness
of the atmospheric emission simulations and to obtain the CXB spectrum
in broader energy range.

\acknowledgements  We are grateful to the referee, Dr. Keith
Jahoda, for the very detailed and helpful comments.  Based on
observations with INTEGRAL, an ESA project with instruments and
science data centre funded by ESA member states (especially the PI
countries: Denmark, France, Germany, Italy, Switzerland, Spain), Czech
Republic and Poland, and with the participation of Russia and the USA.
This research has been partly supported by the Russian Academy of
Sciences programs P-04 and OFN-17, by the Italian Space Agency
contract I/R/046/04 ASI/IASF and Istituto Nazionale di Astrofisica
(INAF).  JMMH and AD funded by Spanish MEC ESP2005-07714-C03-03 grant.

\end{document}